# True polar wander driven by late-stage volcanism and the distribution of paleopolar deposits on Mars

Edwin S. Kite[1], Isamu Matsuyama[1, 2], Michael Manga[1], J. Taylor Perron[3], Jerry X. Mitrovica[4]

1. Earth and Planetary Science, 307 McCone Hall #4767, University of California, Berkeley, CA 94720 USA (kite@berkeley.edu)
2. Department of Terrestrial Magnetism, Carnegie Institution of Washington, Washington, DC.
3. Earth, Atmospheric and Planetary Sciences, MIT, 77 Massachusetts Ave., Cambridge, MA.
4. Department of Physics, University of Toronto, 60 St. George Street, Toronto, Ontario, Canada.

**Abstract.**

The areal centroids of the youngest polar deposits on Mars are offset from those of adjacent paleopolar deposits by 5-10°. We test the hypothesis that the offset is the result of true polar wander (TPW), the motion of the solid surface with respect to the spin axis, caused by a mass redistribution within or on the surface of Mars. In particular, we consider the possibility that TPW is driven by late-stage volcanism during the late Hesperian to Amazonian. There is observational and qualitative support for this hypothesis: in both North and South, observed offsets lie close to a great circle 90° from Tharsis, as expected for polar wander after Tharsis formed. We calculate the magnitude and direction of TPW produced by mapped late-stage lavas for a range of lithospheric thicknesses, lava thicknesses, eruption histories, and prior polar wander events. We find that if Tharsis formed close to the equator, the stabilizing effect of a fossil rotational bulge located close to the equator leads to predicted TPW of <2°, too small to account for observed offsets. If, however, Tharsis formed far from the equator, late-stage TPW driven by low-latitude, late-stage volcanism would be 6-33°, similar to that inferred from the location of paleopolar deposits. A mass of $4.4 \pm 1.3 \times 10^{19}$ kg of young erupted lava can account for the offset of the Dorsa Argentea Formation from the present-day south rotation pole. This mass is consistent with prior mapping-based estimates and would imply a mass release of $CO_2$ by volcanic degassing similar to that in the atmosphere at the present time. The South Polar Layered Deposits are offset from the present rotation pole in a direction that is opposite to the other paleopolar deposits. This can be explained by either a sequential eruption of late-stage lavas, or an additional contribution from a plume beneath Elysium. We predict that significant volcanic activity occurred during the time interval represented by the Basal Unit/Planum Boreum unconformity; Planum Boreum postdates the Promethei Lingula Lobe; and that the north polar deposits span a substantial fraction of Solar System history. If the additional contribution to TPW from plumes is small, then we would also predict that Tharsis Montes formation postdates the the Promethei Lingula Lobe of the South Polar Layered Deposits. We conclude with a list of observational tests of the TPW hypothesis.

## 1. Introduction.

Paleopolar deposits in the polar regions of Mars are not centered on the present spin axis. Although obliquity changes are probably responsible for the waxing and waning of ice-rich deposits in the Martian low latitudes (Forget et al., 2006) and mid-latitudes (Head et al., 2006), it has proved problematic to account for these near-polar offsets using obliquity change. Moderate increases in mean obliquity (to ~35° from the present ~25°) extend the latitudinal range over which water ice is stable, but this expansion should be approximately symmetric about the pole - in contrast with the grossly asymmetric distribution of the paleopolar deposits. At still higher obliquity (≥40°), all Mars General Circulation Models (GCMs) predict fast loss of polar water to form a low-latitude ice necklace (Mischna et al., 2003).

Topographic control of atmospheric circulation and ice mass balance has been proposed as an explanation for the offset of some present-day polar deposits. The low topography of the Hellas and Argyre impact basins forces a planetary wave which accounts for the offset of the carbon-dioxide Southern Residual Cap (SRC) from the present-day spin axis (Colaprete et al., 2005). The thickest part of the southern polar layered plateau underlies the SRC, consistent with topographic control of long-term water-ice deposition as well. However, Hellas and Argyre are both Noachian (> 3.6 Ga), and the only significant topographic changes in the Southern Hemisphere since the mid-Hesperian (~3.4 Ga), in the Tharsis and Syria provinces, could have little effect on either the magnitude or direction of topographically-controlled asymmetry (Colaprete et al., 2005). In addition, we see a larger pattern involving offset of both the north and south polar deposits that suggests that the location of the water ice caps and associated ancient deposits are not due to asymmetric atmospheric effects.

We now turn to the alternative hypothesis of late-stage true polar wander, first proposed for the Martian polar deposits by Murray & Malin (1973). True Polar Wander (TPW) is the movement of a planet's entire lithosphere with respect to the spin axis in response to changes in mass distribution. Here we apply TPW theory (Gold, 1955; Goldreich & Toomre, 1969; Willeman, 1984; Matsuyama et al., 2006) to assess whether relatively recent volcanism in the Martian low- and mid-latitudes could be responsible, through surface-load driven TPW, for shifts in the distribution of Mars' polar ice sheets since ~3.5 Ga. Because most Martian volcanic and tectonic activity predates ~3.5 Gya, we refer to this as “late-stage” TPW.

The basic idea is shown in Figure 1. Lavas postdating the older polar deposits on Mars, but predating younger polar deposits, change the geoid and so can shift the spin axis and the location of ice accumulation. Late-stage TPW on Mars is a long-standing hypothesis (Murray & Malin, 1973; Schultz & Lutz, 1988; Tanaka, 2000; Fishbaugh & Head, 2000), but both TPW theory, and Martian geological and geodetic data, have been refined in recent years and so permit a quantitative analysis. Both surface and internal loads can drive TPW, but young volcanic loads are more amenable to testing as a possible driving load because their size and location are relatively well constrained. Our intent is to test if these geologically plausible loads can produce the required TPW to explain the polar deposit offsets.

True Polar Wander of modest magnitude (≤10°) has been previously suggested to be responsible for the distribution of both north (Murray & Malin, 1973; Fishbaugh and Head, 2001) and south (Tanaka, 2000) polar materials. Because the oldest polar materials are Early Amazonian in the north and Late Hesperian in the south, late-stage TPW requires loading the lithosphere in the Late Hesperian or Amazonian. By this time the bulk of the Tharsis load, which dominates the Martian geoid, was already in place (Phillips et al., 2001). However, volcanism has persisted throughout the Amazonian, resurfacing much of the Tharsis and Elysium regions – 15% of the planet's surface, of which some four-fifths lies north of the present-day equator (Nimmo and Tanaka, 2005). The areal centroid of these lavas is 21N 215E; without other loads they would have driven the pole north along the 215E line. This is approximately the direction of offset between polar and paleopolar deposits in the north – an observation that motivates our more detailed study.

Here we evaluate the conditions under which TPW driven by recent surface loading may account for the offset between past and present polar caps. Our scenario complements, but does not require, proposed TPW events of greater (≥30°) magnitude and age that have been invoked to solve geological puzzles earlier in Martian history (e.g. Perron et al., 2007).

## 2. The geologic record of polar ice sheets

The residual ice caps at the Martian poles (Herkenhoff et al., 2006) are underlain and abutted by materials that have been mapped as the eroded remnant of past ice-sheet deposits (Figure 2). In the north, relatively young polar layered deposits (Planum Boreum units of Tanaka et al., 2008), compositionally dominated by water ice and ~1500m thick, are underlain by older, more lithic-rich materials (the Basal Unit of Fishbaugh and Head, 2005, platy unit of Byrne and Murray, 2002, lower unit of Edgett et al., 2003, Rupes Tenuis and Planum Boreum cavi units of Tanaka et al., 2008; we use the first term), comprising ≤1400m of interbedded layers of sand and fractured ice. Radar sounding (Phillips et al., 2008) has confirmed the indication from geological mapping (e.g. Fishbaugh & Head, 2005) that the center of these older, more lithic-rich materials is offset from the center of Planum Boreum by ~300km (~5°) along 180E (Figure 2). Near 180E, these sand-rich materials embay and overlie the Scandia terrain, which bears large rimmed depressions interpreted as dead-ice topography (Fishbaugh and Head, 2000). To generate these features would require thick (> 100m) and extensive (> $5 \times 10^4$ $km^2$) 'ice sheets', by contrast with the thin (≤10 m) midlatitude icy layers associated with the recent high-obliquity ice ages (Head et al., 2003).  It is unclear how far Scandia morphologies extend beneath the Basal Unit. If they mark the perimeter of a past ice sheet of similar radius to that of the Planum Boreum plateau, then its center was offset from the center of Planum Boreum by ~450km (an additional ~150km) along the 200E line. All of the above units are superposed on the interior Vastitas Borealis plain, which has an earliest Amazonian crater-retention age by definition (Tanaka, 2005).

Still larger offsets are inferred in the south. The $CO_2$-dominated Southern Residual Cap is offset by 2.7° from the present-day spin axis, and the distribution of the underlying, water-ice Planum Australe plateau is very asymmetric. A lobe, wedge-shaped in cross-section (Plaut et al., 2007b), extends in the 180E direction, so that the plateau's areal centroid is at ~85S, although its volumetric centroid is closer to the pole (Figure 2). Stratigraphic correlation shows that this lobe represents a subset of Planum Australe's history (the Promethei Lingula Layer sequence of Milkovich & Plaut, 2008). Older and younger sequences within Planum Australe are only found poleward of 80N; this may reflect their original extent, or asymmetric aeolian erosion (e.g., Koutnik et al., 2005). Limning Planum Australe between longitudes 240E and 105E, and most extensive near 300E and near 10E, is a region of sinuous, braided ridges and smooth plains punctured by sharp-rimmed depressions, which is mapped as the Dorsa Argentea Formation (DAF) (Fig. 6b). Radar shows an ice-rich layer extending to 600-900m depth in this area (Plaut et al., 2007a), supporting the subglacial-melting hypothesis for the DAF (Head and Pratt, 2001), which has a Hesperian crater-retention age. Radar shows kilometer-deep depressions beneath Planum Australe resembling the Cavi Angusti within the DAF (Plaut et al., 2007b), so the DAF likely continues beneath the polar plateau. This interpretation, indicating a paleo-cap offset by ~10° (~600km) from the spin axis along the 335E line, is also supported by the pattern of outcrop.

**3. Reorientation of Mars by surface loading**

A rotating fluid planet in hydrostatic equilibrium, subject to a superimposed nonequatorial surface load, reorients to place that mass excess on the equator. However, if a planet has a lithosphere with elastic strength at spherical harmonic degree two, the equilibrium location of the surface load is not at the equator, but at an intermediate latitude less than that of loading (Willemann, 1984; Ojakangas and Stevenson, 1989; Matsuyama et al., 2006, 2007).

On Mars, late-stage TPW is also affected by 1) the strong tendency of the preexisting Tharsis load to remain near the equator (Perron et al., 2007) (all plausible young loads are much smaller than Tharsis), and 2) the unknown orientation of a fossil equatorial bulge, a “frozen-in” relic of the planet's spin orientation prior to Tharsis emplacement (Willemann, 1984). Geoid constraints favor solutions with Tharsis forming close to the equator – in which case the bulge would also lie close to the present-day equator – but still permit Tharsis to have formed far from the equator, in which case the bulge would lie at a high angle to today's equator (Daradich et al., 2008). We treat both configurations.

We follow Matsuyama et al. (2006) to calculate the inertia tensor perturbations and the corresponding pole position by diagonalizing the inertia tensor. First, we consider the perturbations associated with the young loads, Tharsis, and the remnant bulge to find the spin-axis orientation prior to the formation of Tharsis and the young loads. Then, given this initial spin-axis orientation, we consider the perturbations associated with Tharsis and the remnant bulge alone to calculate the spin-axis orientation after the formation of Tharsis (Table 1a). We use a standard value for the size of the geoid perturbation due to the Tharsis load (from Willemann, 1984), although the true value may be greater or smaller.

In practice, Martian volcanism is episodic (Wilson et al., 2001). The viscous relaxation time of Mars determines whether TPW would also have been episodic, or smoothed. We simplify by considering the equilibrium response of the planet to a single loading event – equivalent to assuming that the viscous relaxation time is much shorter than the interval separating younger from older polar deposits.

## 4. Calculation of TPW due to young volcanic loads

We focus on igneous loads emplaced on top of the elastic lithosphere ("surface volcanic loads") – we justify this in §6.3). Using geological unit boundaries slightly modified from digitally renovated global maps (Skinner et al., 2006; Scott and Tanaka, 1986; Greeley and Guest, 1987), we assess the total thickness of young volcanic materials (lavas, plus relatively shallow intrusions) for each unit (Figure 3; Table S1).

Some units are smooth plains, others are edifices (lobes and cones). We assume uniform thicknesses for plains lavas, since sub-plains topography is rarely known. Our values resemble those of Greeley and Schneid (1991). For edifices, we either assume uniform thickness (a late-stage drape over a Noachian-Early Hesperian core), or truncate near the base (assuming that the majority of the edifice is late-stage). Planar truncation does not account for the volume of the filled-in flexural pit beneath the edifice, so the truncated volume underestimates the total edifice volume. There is disagreement over the volume-averaged age of volcanoes that have very young surface ages. For example, Dohm et al. (2001) reconstruct Olympus Mons as forming in the Late Hesperian/Amazonian, but Werner (2005) favors formation of Olympus in the Noachian/Early Hesperian. Elastic thickness estimates entail geothermal gradients generally consistent with late formation (McGovern et al., 2004). However, the extent to which these estimates are quantitatively reliable is unclear (Belleguic et al., 2005), especially if volcanoes form above hotspots. To span this uncertainty, we assign 'small', 'medium' and 'large' thicknesses to each unit (Figure 3; Table S1). We also model TPW driven by drapes of uniform thickness within the boundary of Late Hesperian/Amazonian volcanic materials given by Nimmo and Tanaka (2005).

[New age estimates and volumes became available as this paper was in revision. Using crater-counts on High-Resolution Stereo Camera (HRSC) images, Werner (in press) infer that the bulk of the Tharsis Montes, Olympus Mons, and Elysium were emplaced before the Late Hesperian, although flank volcanism, caldera activity, and extrusion of flood lavas continued through to the Late Amazonian. These results suggest that Late Hesperian/Amazonian volcanism can be approximated as a uniform drape of 100 m to perhaps a few hundred meters thickness, similar to Figure 3a.]

Load density $\rho$ is set to $3\text{x}10^3$ kg m$^{-3}$. This is conservative for the largest volcanoes (Belleguic et al., 2005), but overestimates the density of the Medusae Fossae Formation ($\rho \leq 1.9 \text{ x } 10^3$ kg m$^{-3}$; Watters et al., 2007). This error is small compared to the range in thickness estimates. Only that portion of the load that is uncompensated will drive TPW. For each load, we self-consistently calculate flexure using fluid Love numbers, and obtain the total geoid perturbation due to the load.

We model late-stage TPW for a range of elastic thicknesses ($T_e$), for both the cases where Tharsis formed close to and far from the equator (Table 1a). Because, once formed, Tharsis must remain near the equator, the direction of TPW is always close to 160E (90° from the center of Tharsis). For fixed $T_e$, the magnitude of TPW is approximately linear in load volume for reasonably-sized loads.

However, it is much greater for the case where Tharsis forms far from the equator than where Tharsis forms close to the equator; even with the `large' load, we obtain no more than 1.6° of TPW in the latter case. This is because the configuration with the fossil rotational bulge at a small inclination to the present-day equatorial plane is very stable (Perron et al., 2007). For fixed load volume, the extent of late-stage TPW increases with increasing $T_e$, because the level of load compensation decreases; this effect dominates the increasing magnitude of the remnant rotational bulge, which acts to resist TPW. Results for drapes of uniform thickness behaved similarly, implying that it is the magnitude of young volcanism, rather than its exact distribution, that is more important in determining the TPW response. For simplicity, we assume that the same $T_e$ characterizes the remnant rotational bulge and the young volcanic loads.

To illustrate the relationship between volcano position and TPW response, we also calculated the response to the `large' estimates of volume for individual volcanic provinces (Table 1b). Polar offset vectors so calculated cannot be meaningfully added, because the TPW response is non-linear. Apart from Alba Patera, whose effect is to increase the Tharsis load and move Tharsis slightly closer to the equator, most paleopoles lie along the great circle 90° from Tharsis (Figure 5). The sense of the shift induced by these volcanic provinces can be understood by inspection of Figure 4. The Medusae Fossae Formation and Tharsis Montes provinces have volumetric centroids S of the center of Tharsis in the west-of-Tharsis hemisphere, giving polar shifts in the opposite sense to that of the most voluminous provinces, which lie N of the equator in the west-of-Tharsis hemisphere. Among these voluminous provinces, TPW magnitude is strongly dependent on load longitude (Matsuyama et al., 2006). Elysium is only slightly more massive than Olympus Mons, but because its longitude is close to 90° from Tharsis it drives a shift ~3 times greater than Olympus Mons.

Stratigraphic correlations between Elysium and Tharsis are imprecise, although widespread volcanism on the Elysium rise must have concluded during the Early Amazonian (Tanaka et al., 1992).

The consensus among stratigraphers is that overlapping periods of activity occurred at Mars' main late-stage volcanic centers, rather than sequential eruption at one province followed by another. For all these reasons, a good starting point when discussing polar wander paths is to assume that the activities of the Tharsis and Elysium volcanic centers were proportional. However, as will be discussed in §5, our TPW results illuminate the stratigraphic relations between volcanism and polar activity, allowing us to reconstruct an alternative 'sequential' history (provided that TPW is in fact responsible for the observed offsets). This sequence corresponds to the timing of near-surface emplacement of the bulk of the volume of the volcanoes, not to emplacement of the youngest crater-bearing surfaces.

## 5. Comparison with paleopoles inferred from geology

### 5.1. Geologically inferred paleopoles

Areas of past water ice concentration have been inferred using geology (Tanaka, 2005), and neutron-spectrometer maps of upper-metre H abundance (Dohm et al., 2007). Although the second dataset may contain valuable information about ancient (>> 5 Mya) ice sheets, the possibility that H is present in salts rather than ice (Basilevsky et al., 2006), and overprinting by recent (≤ 5 Mya) atmosphere-regolith exchange (Jakosky et al., 2005), complicates interpretation of these shallow materials. We define paleopolar deposits using topographic and digitized geological data only (Figure 6): Tanaka and Scott (1987) for the south; Tanaka et al. (2005) for the north. The Scandia formation is mapped (Tanaka et al., 2005) as being much more extensive than the area of large, rimmed depressions (Scandia Cavi). Rather than use the entire Scandia formation to define a paleopole, we use Mars Orbiter Laser Altimeter (MOLA) topography to draw the margins of the Scandia depressions and eskers. A zone of unusual, large moated domes (Scandia Tholi) encircles the Scandia depressions. These may have been built up by the extrusion of shallow slurries (Tanaka et al., 2003), and slurry extrusion could occur without differential loading from an advancing or retreating ice sheet (Kite et al., 2007). But the domes have also been suggested to represent kame-and-kettle topography (Fishbaugh & Head, 2000), so we use MOLA topography to define their convex hull as a possible paleopolar deposit. We define a single paleopole using the Olympia dome, which is the only extensive area where the Basal Unit probably lies close to the surface. This involves the assumption that the part of the Basal Unit making up the bulk of the Olympia dome is conformable on, or contemporaneous with, the part of the Basal Unit that forms the base of Planum Boreum (Fishbaugh & Head, 2005); however, Tanaka et al. (2008) argue that the first part is separated from the second by a major unconformity.

Polar caps are centered close to the spin axis today (average deviation 2.4°, Table 2), and are likely to have tracked the spin axis in the past, at least during conditions of low (<40°) obliquity. We show the distance between young polar caps and the present-day spin axis in Figure 6, by dashed lines bracketing the calculated polar wander track. This is our estimate of the 'proxy error' involved in using polar caps as proxies for the location of the spin axis. Radar indicates that the Basal Unit continues under much of Planum Boreum (Putzig et al., 2008), and the spatial distribution of Dorsa Argentea Formation outcrop strongly suggests that it continues under the thickest part of Planum Australe. Therefore, the areal centroid of present-day outcrop of a paleopolar deposit cannot be considered a paleopole. Instead, we approximate past ice sheets by circles, and define a simple penalty function to evaluate the goodness of fit of a given circular planform to geological boundaries. Incorrect inclusion of non-paleopolar material and exclusion of paleopolar material both score +1 per grid cell (~ 9km$^2$), and correct inclusion of paleopolar material is rewarded with a score of -1 per cell. Overlying materials are masked out and not scored. We then define the paleopole for a given geological unit to be the center of the circular ice-sheet with the minimum score, searching systematically over cap latitude and longitude. The circle that best fits Olympia Planum matches the area of Basal Unit independently mapped by radar (Putzig et al., 2008), lending credence to our assumption that paleopolar caps were approximately circular. We do not attempt a formal error estimate for the paleocenter locations: this would have to take into account incomplete geological preservation, mapping imprecision, non-circular ice sheets, and possible asymmetric retreat. We consider that these effects introduce an error in assigning a centroid location to a paleopolar deposit comparable to the 'proxy error' defined above. This additional 'measurement error' is shown by the outer green circles in Figure 6.

In the north, we only consider past ice-sheets with radii equal to that of present-day Planum Boreum (Greenwood et al., 2008). We find best-fitting pole locations for the Olympia dome, for the Scandia depressions, and for the Scandia depressions and mounds taken together (Table 1a). Ice outliers, large areas of perennially high albedo south of the Olympia dome, bracket the polar wander path associated with progressive emplacement of a volcanic load (§4; Figure 6a). They have only superficial topographic expression, and their age is uncertain. They may be recent deposits unrelated to the Scandia depressions, in which case it would be inappropriate to fit them. Alternatively, if they are outliers of a formerly more extensive polar plateau (Zuber et al., 1998), then they should be weighted when defining paleopoles. We plot only the first possibility, which we consider more likely because the ice outliers superpose most other deposits (Table 1a; Figure 6) and are still annually frosting and defrosting (Calvin and Titus, 2008). The calculated polar wander path including the ice outliers (not shown) runs along 180(±10)E.

Taken together, our paleopoles in the north define a polar wander path of magnitude (5±2)° (Figure 6a). This increases to (7±2)° if the enigmatic Scandia mounds also indicate ice-sheet processes. The Basal Unit paleopole is almost as close to the spin axis as is the upper part of Planum Boreum, so TPW is only one of several possible explanations for the offset of the Basal Unit from the spin axis and from the center of the upper part of Planum Boreum. For example, pre-existing (and now-buried) topography could have seeded growth of an ice-sheet. Subsequently, the Basal Unit topographic rise might have shifted atmospheric circulation to favor off-center deposition of younger volatile-rich deposits.

In the south, caps with radius equal to that of Planum Australe cannot give a good fit to the location of the DAF. When we allow the paleo-ice-sheet radius to increase, we obtain a much more satisfactory fit. This may be because the Martian surface lost water between DAF time and Scandia time (Kulikov et al., 2007), so that the DAF ice-sheet was more voluminous. In addition, higher lithospheric heat flux (Solomon et al., 2005) may have permitted the ice-sheet to flow downslope, leading to a noncircular planform and greater width:height ratio. Geological syntheses based on image interpretation (Ghatan & Head, 2004) suggest that the DAF ice-sheet was at least locally hydrologically active, indicating a bed temperature near the pressure melting point. This would have permitted much more rapid ice creep than within the present-day polar layered deposits (Goldsby & Kohlstedt, 2001). Bedrock topography in the south polar region favors asymmetric accumulation and flow: it is more rugged, and has steeper long-baseline tilts, than the north polar region (Plaut et al., 2007b). The best-fitting ice sheet, with radius 1020 km (q.v. 700 km for Planum Australe), is centered along our calculated TPW path. Its offset relative to the present-day spin axis is 7.9°, and we use this value in the Discussion. However, given the importance of Southern Hemisphere topography in controlling precipitation patterns near the south pole, it is arguably better to measure the offset relative to the SRC (5.2°). In this case, the volume of volcanism needed to generate TPW is reduced by 1/3, but our conclusions are not qualitatively affected.

**5.2 Assessment of the late-stage TPW hypothesis**

In Figure 6 we compare paleopole locations inferred from geological constraints (Table 1a) to those computed using loading theory (Table 1a). As episodic volcanism progressively emplaced the young volcanic load, the magnitude of TPW would have increased. Therefore older paleopolar deposits should show progressively larger offsets from the present-day spin axis, a prediction that matches observations (Figure 6) and correlations between north and south (Figure 7).

If Tharsis formed close to the equator, volcanically-driven TPW can explain at most the location of the Basal Unit, even with the ‘large’ estimated volcanic load. Accounting for the older deposits requires instead that Tharsis formed far from the equator. In that case, even the 'small' load gives TPW sufficient to explain the offset between the present-day spin axis and the DAF.

The trend of the paleopoles in the north clearly deviates from the great circle 90° from Tharsis (Figure 6a). However, Paleopoles 1 – 2 (which lie close to the great circle 90° from Tharsis) are more reliable than Paleopole 3 (which does not) because of the greater area of outcrop of the Planum Boreum and the Basal Unit compared to that of the Scandia depressions. Fitting caps to relatively small outcrops introduces additional uncertainty into the pole calculation. In turn, Paleopoles 1-3 are much more reliable than  Paleopole 4 (the biggest outlier), because we do not know what created the Scandia Tholi. Moreover, the ice outliers, which are thickest in the 110E–180E sector, might mute topographic evidence for ice-sheet action to the west of our paleopoles, introducing an eastward bias in paleopole longitude. Therefore, we consider the distribution of paleopolar deposits in the north polar region consistent with TPW, although certainly not compelling when considered in isolation.

In the south, all geologically-inferred paleopoles lie along our calculated polar wander path (Figure 6b). As expected given its greater age, the DAF shows a larger offset from the present-day spin axis than does the Basal Unit. The center of Planum Australe lies along our calculated TPW path, but in the opposite direction from that expected for proportional emplacement of young volcanic loads (§5.3).

Given the plausibility of our input parameters, and the acceptable match between observed and calculated polar wander magnitudes and directions, we favor the late-stage TPW explanation for the distribution of paleopolar deposits on Mars. However, if later research confirms that Tharsis in fact formed close to the equator, this will rule out TPW driven by surface loads as an explanation for the distribution – unless late-stage Martian magmatism was much more voluminous than in current models. Gravity data to degree and order 2 are better fit by a small-TPW scenario than by a large-TPW scenario (Daradich et al., 2008), but that work did not consider loads other than Tharsis and the fossil rotational bulge (such as internal loads, or the late-stage surface loads considered in this paper). More detailed modeling of gravity anomalies and geodynamic history might help to constrain the latitude at which Tharsis formed.

**5.3. Example integrated volcanic and polar history of Late Hesperian/Amazonian Mars**

Because Tharsis dictates the path of late-stage TPW, our inferred TPW path cannot be used to infer the exact location of its causative loads (Figure 4). For example, the difference in polar wander direction for TPW driven by Elysium versus Olympus is negligible, despite the >70° of longitude separating Elysium and Olympus. The longitude resolution of our geologically inferred paleopoles is also poor, so we cannot constrain load longitude much better than by quadrant (Figure 4). However, in combination with upper limits on load volume from MOLA topography, and lower limits on unit age from crater counts, we identify a possible sequence of volcanic loading episodes and pole shifts, and polar deposition episodes (Figure 7).

The sequence starts with deposition of the Dorsa Argentea Formation in the Late Hesperian. The Elysium rise is already partly in place (Figure 7a). Waning Elysium activity, increasingly supplemented by proto-Olympus Mons, drives the pole ~5.5° towards 335E (towards the hemisphere centered on the Martian meridian) during Scandia and prior to Basal Unit time (Figure 7b). The top of the Basal Unit marks a major unconformity in the north. Olympus Mons and Arcadia flood lavas move the pole a further ~8.1° meridian-ward, ~5.7° beyond its present location. The Promethei Lingula Lobe of the South Polar Layered Deposits forms after this loading event (Figure 7c). Next, the locus of volcanism shifts east to the Tharsis Montes, reversing the sign of TPW and bringing the spin axis to its present configuration. Thicker lithosphere allows a given load to produce a greater geoid anomaly, so cooling of the planet enhances TPW during this final stage. Planum Boreum is deposited during this TPW event, a ~5.7° shift in the anti-meridian direction (Figure 7d), accounting for its offset from the present-day north rotation pole (Figure 6a). The Medusae Fossae Formation, Syria and Alba play subsidiary roles (Figure 5).

This sequence of events assumes there has been little change in the contribution to the geoid from internal loads and dynamic topography over the past ~3.5 Gyr. In §6.3, we relax this assumption, and suggest that construction of the Elysium province, followed by cessation of plume activity beneath Elysium, is an alternative and perhaps simpler explanation of the observations. If late-stage volcanism at the Tharsis Montes has been volumetrically minor (Werner, in press), late-stage Tharsis Montes volcanism would have been insufficient to generate the ~5.7° shift in the anti-meridian direction shown in Figure 7d, and a contribution from internal loads would be required to explain this step of Mars' inferred TPW history.

## 6. Discussion

### 6.1. Implications for rates of melting and degassing

$T_e$ is probably ≥ 300 km today (Phillips et al., 2008), but was 93±40 km at the time of Olympus Mons emplacement (Belleguic et al., 2005). Given the decay of the mantle's radiogenic heat source, the relevant $T_e$ is probably close to ~100 km, the value for the Early Amazonian when most of the load would have been emplaced.

Given the ~linear relationship between load volume and magnitude of TPW (Table 1a) for geologically defensible loads, we can use geologically inferred TPW to estimate the subsequent volume of volcanism (Figure 8). For example, taking the Dorsa Argentea Formation to be ~3.6 Ga old (Hartmann, 2005), the average rate of volcanism since is (4.1±1.2) x $10^{-3}$ $km^3$ $yr^{-1}$. This is less than the pre-MOLA estimate of Greeley and Schneid (1991), 7.3 – 14.6 x $10^{-3}$ $km^3$ $yr^{-1}$, but matches thermal evolution models (O'Neill et al., 2007).

With an estimate of the $CO_2$ content of Martian magma – 516 mg/kg (O'Neill et al., 2007), with large uncertainties (Hirschmann and Withers, 2008) – we can estimate total late-stage volatile release to the atmosphere. For the case where Tharsis forms far from the equator, and $T_e$ = 100km, (4.4±1.3) x$10^{19}$ kg magma (which, if completely degassed, would have released 5.8±1.8 mbar $CO_2$) should have been extruded since DAF time, and (2.8±1.3) x$10^{19}$ kg magma (which, if completely degassed, would have released 3.6±1.7 mbar $CO_2$) since the Scandia depressions formed. This falls to 3.3±1.0 mbar and 2.1±1.0 mbar respectively if with $T_e$ = 200 km. Quoted errors consider only the error in paleopole positions, not in $T_e$. Our best-estimate of total late stage degassing, ~6 mbar, is also the size of the present-day near-surface $CO_2$ reservoir (6.4 mbar in the atmosphere, and 0.4 mbar in the SRC).

Present-day Martian atmospheric pressure reflects the balance of Late Hesperian/Amazonian volcanic degassing and atmospheric loss, added to whatever atmosphere was inherited from 'early Mars'. The inherited atmosphere would have been very tenuous if the active young Sun drove $CO_2$ loss rates greatly in excess of volcanic degassing rates during the Noachian/Early Hesperian (Manning et al., 2006). In that case, the rough equivalence of TPW-derived volcanic degassing estimates and present-day atmospheric pressure would require that atmospheric loss rates were smaller than volcanic degassing rates, integrated over the Late Hesperian/Amazonian. In other words, long-term average atmospheric mass and globally-averaged surface temperature may have been *increasing* over the last three-quarters of Martian history, with important implications for geomorphology and the likelihood of surface liquid water (see also Richardson & Mischna, 2005). The requirement that atmospheric loss rates be small is consistent with the paucity of carbonates on the Martian surface (Bibring et al., 2006), and with Mars Express measurements of present-day atmospheric escape rates – which are equivalent to only 0.2 – 4 mbar $CO_2$ extrapolated over 3.5 Ga (although not all processes have been quantified) (Barabash et al., 2007).

### 6.2. Are Planum Boreum and Planum Australe old enough to have experienced TPW?

Our proposed TPW scenario implies an age for Planum Australe and the upper part of Planum Boreum that is older than some estimated ages, but consistent with others. If they formed recently (e.g., <5 Myr in the North; Phillips et al., 2008), geologically detectable TPW is unlikely to have occurred since. TPW acts to filter out high temporal frequencies, and the only non-polar changes in surface load since 5 Mya have been the emplacement of thin (Campbell et al., 2007) lavas, perhaps the uppermost member of the Medusae Fossae Formation (Watters et al., 2007), and perhaps partial sublimation of non-polar icy deposits (Head et al., 2006). Collectively, these load changes could have produced only minor TPW. However, estimating the age of icy deposits is difficult, because sublimation of overburden may remove craters, and crater topography possibly undergoes viscous relaxation. The largest craters, though few in number, imply a surface age of >330 Myr for the upper part of Planum Boreum, and >500 Myr for Planum Australe (Pathare et al., 2005). Because the stratigraphy of the north polar layered deposits cannot yet be definitively linked with a known time scale (Perron and Huybers, 2009), it remains possible that the upper part of Planum Boreum is significantly older than a few million years. Therefore, either or both Planum Australe and the upper part of Planum Boreum may be old enough to have undergone geologically measurable TPW, which may help to explain the offset of their areal centers from the present-day spin axis.

### 6.3. Other loads

If mantle plumes underlie Tharsis and Elysium, they may have changed intensity (O'Neill et al., 2007) or position over time. However, the geoid anomaly at degree 2 (the wavelength relevant to TPW) produced by a plume of constant flux and position changes little as the lithosphere thickens over time (Roberts and Zhong, 2004). Rouby et al. (2008) show that plume initiation could have led to 1-15° of TPW on pre-Tharsis Mars, setting aside stabilization by the fossil rotational bulge.

Extensive volcanism on the Elysium rise ceased in the Early Amazonian (Tanaka et al., 1992). The decline of volcanic activity may correspond to a waning of plume activity beneath Elysium. If so, then the 'overshoot and rebound' relative to today's spin-axis recorded by the locations of the polar deposits could be due solely to Elysium. Our reasoning is as follows. On Mars, the geoid anomaly due to mantle plumes at wavelengths appropriate for Elysium is probably positive (upwards), though small (Roberts and Zhong, 2004); the positive anomaly due to dynamic topography supported by the rising mantle exceeds the negative anomaly due to the low-density mantle forming the plume. Therefore, with plume-supported volcanism, the combination of a dynamic, internal load and a growing surface load induces greater TPW than the surface load alone. This could correspond to the evolution from a) to c) in Figure 7 (the 'overshoot') . Removal of the internal load, for example by a waning plume, is in effect a negative load. This could correspond to the evolution from c) to d) in Figure 7 (the 'rebound').

Polar wander driven by dust, ice, or material deposited by the waning circum-Chryse outflows was probably small compared to the volcanically-driven wander assessed here. The thin Vastitas Borealis outflow channel deposits certainly predate the Scandia terrain, and probably largely predate the Dorsa Argentea Formation. The TPW contribution of the ice sheets themselves would have been minor unless vastly more ice was available for atmospheric transport in the past than at present.

On Earth, basalt underplating the low-density crust during plume impingement may match or exceed erupted volumes (Cox, 1995). Such 'bottom loading' on Mars would complicate our analysis.

We cannot distinguish between true polar wander of the lithosphere-plus-mantle relative to the spin axis, and differential motion of the lithosphere with respect to the mantle. Our calculations in §3 assume the former. The alternative has recently been proposed (Zhong, 2009), and would lead to very different implications for Martian volcanic (and climatic?) history.

**6.4 Further tests**

We have established the plausibility of geologically significant late-stage TPW on Mars, but our conclusions depend on assumptions about the volcanic and polar history of Mars that need to be tested. The TPW hypothesis would be disfavored if analysis of crater size-frequency distributions showed a Noachian/Early Hesperian density of buried craters at shallow depths throughout the area of young volcanism. Provided that the intrusive:extrusive volume ratio is comparable to terrestrial values (~5:1; White et al., 2006), this could bound the volume of late-stage near-surface loads as less than required to drive the proposed TPW. On the other hand, tectonic mapping might reveal patterns of fault reactivation parsimoniously explained by lithospheric stresses due to late-stage TPW (Melosh, 1980). Our best-fitting Scandia paleopoles imply that features similar to the Scandia rimmed depressions extend under Planum Boreum about as far as the present-day spin axis. This predicts that the basal interface of the Basal Unit should have hundreds of meters of relief at 10km – 100km scales, contrasting with the low ~10km-scale roughness of the basal interface of Gemini Lingula (Phillips et al., 2008) and of Vastitas Borealis (Kreslavsky & Head, 2000). If radar shows these features to be absent, asymmetric retreat of a formerly more extensive ice sheet (Fishbaugh & Head, 2001) may be a better explanation for the Scandia features.

Our late-stage TPW scenario alters the topographic boundary conditions for Noachian/Early Hesperian climate by 1) removing volcanic edifices, and 2) shifting zonal climatic belts relative to the lithosphere. Ventifacts at the mid-Hesperian Mars Pathfinder (MPF) landing site are oriented oblique to the strongest present-day winds (Greeley et al., 2000); this cannot be accounted for by obliquity change (Fenton & Richardson, 2001; Haberle et al., 2003). If a GCM using the altered topographic boundary conditions provided by our TPW scenario yields strong winds at the MPF site which parallel the observed ventifacts, the hypothesis of late-stage TPW would be supported.

Also, if the significant offset between the centers of the Basal Unit and the upper NPLD is due to TPW, the Basal Unit must predate a substantial fraction of Late Hesperian / Amazonian volcanism.

Given that a recent surge in volcanism is unlikely (but not impossible), this entails a mid-Amazonian or earlier Basal Unit age. Consequently, we predict that the north polar deposits span a substantial fraction of Solar System history. If Martian polar stratigraphy records orbitally forced variations in insolation (Montmessin, 2006), then the polar deposits of Mars might greatly extend the time range over which geology constrains orbital dynamics (Olsen and Kent, 1999).

## 7. Conclusions

Late Hesperian/Early Amazonian paleopolar deposits near the Martian poles are offset from the present-day poles in a manner inconsistent with simple obliquity change. We find that true polar wander driven by late-stage, low-latitude volcanic loads can account for these observations, if (and only if) Tharsis formed far from the equator. Gravity data favor scenarios where Tharsis forms close to the equator, but permit scenarios where Tharsis forms far from the equator (Daradich et al., 2008).

Three of the four paleopolar deposits on Mars can be explained by proportional emplacement of volcanic load, but the South Polar Layered Deposits cannot be explained in this way. The SPLD paleopole can, however, be accounted for if the Tharsis Montes postdate the South Polar Layered Deposits or if internal (plume-related) loads changed with time. By linking widely separated TPW-inducing and pole-tracking deposits, TPW permits a new method of stratigraphic correlation that is independent of crosscutting relationships and crater counts.

Our calculations illustrate (Fig. 7) that the locations of *all* of the paleopolar deposits near the Martian poles could be the result of TPW, but our confidence level is high for only one of the four deposits, the Dorsa Argentea Formation. Younger deposits are closer to the present-day spin axis, and it is possible that meteorological factors are responsible for their offset from the pole. This could be studied with a GCM of sufficient resolution (~1° in latitude).

The hypothesis of true polar wander driven by late-stage volcanism survives our test.

**Table 1.** a) (Northern) paleopoles calculated for TPW driven by young volcanic load. Letters refer to the solutions that are plotted on Figure 6. b) (Northern) paleopoles calculated for TPW driven by individual volcanic provinces. See Figure 3a for province locations. Only `large' loads considered. Far-from-equator solutions are plotted on Figure 5.

**Table 2.** Best-fitting geologically inferred paleopole locations. OD = Olympia Dome; PB = Planum Boreum, upper North Polar Layered Deposits (similar extent to that of residual water-ice cap); SC = Scandia depressions; ST = Scandia mounds; IO = ice outliers; SRC = Southern Residual Cap ($CO_2$ ice); SPLD = South Polar Layered Deposits; DAF = Dorsa Argentea Formation. Deviation of youngest polar deposits from spin axis = (2.4±0.3)°. This uncertainty is one-sided for the SRC because atmospheric effects forced by Noachian topography displace $CO_2$ precipitation patterns meridian-ward. The final column is a (subjective) confidence level that each paleopole is due to TPW. *Requirements for a good geologically inferred paleopole:* Asymmetric retreat of formerly more extensive ice sheets is an alternative explanation for the shape and distribution of Planum Boreum (Fishbaugh & Head, 2001) and Planum Australe. It is possible that asymmetric sublimation and aeolian erosion has affected the shape of the Basal Unit plateau (Tanaka et al., 2008) and Planum Australe (Milkovich & Plaut, 2008) to the extent that their present-day areal centroids are not close to their past areal centroids. Present outcrop distributions can only be useful as a guide to the past location of the spin axis if either: 1) the ice-sheet is preserved largely intact; 2) original ice-sheet volume is decimated, but the areal extent of the plateau is unaltered; 3) the ice sheet undergoes symmetric retreat; or 4) the ice sheet undergoes asymmetric retreat, but leaves morphological signatures such as moraines, eskers, and kettles that allow its original extent to be reconstructed. Asymmetric retreat of cold-based ice sheets (for which none of these conditions need hold) could confound interpretation of the distribution of Martian paleopolar deposits in terms of TPW. However, the original extent of cold-based high-obliquity ice sheets on the flanks of the Tharsis Montes has been reconstructed using moraines deposited during their incomplete retreat (e.g., Shean et al., 2007).

**Table S1.** Load thicknesses. Unit designations are from I-1802AB. For each unit, we either assume a uniform thickness (a relatively thin late-stage drape over Noachian-Early Hesperian materials), or truncate the edifices near their base (which assumes that the bulk of the unit is Late Hesperian or Amazonian). All units are meters (m). Uniform thicknesses are denoted by numbers without appended parentheses. Parenthetical 'T' denotes edifice truncated at this elevation, and thickness taken to be the pixelwise (¼° x ¼°) thickness above that elevation. Negative thicknesses are not permitted. We modified published geological maps by: 1) assigning ¼° x ¼° areas shown on the map as either 'undivided', 'volcano, relative age unknown' or 'crater ejecta' to the nearest dated unit, 2) subdividing members of the Tharsis Montes formation into Arsia Mons, Pavonis Mons and Ascreaus Mons sub-members, and 3) following Bradley et al. (2002), subdividing the Medusae Fossae Formation by lobe rather than by stratigraphic level

**Figure 1.** Azimuthal equal-area projection centered on 165E, 0N, showing relationship between load, TPW response, and polar deposits. Grid spacing is 30°, in both latitude and longitude. Red tint is area of Late Hesperian – Amazonian volcanics (Nimmo & Tanaka, 2005), shown in more detail in Figure 3. Blue and pink tones are polar and paleopolar deposits, shown in more detail in Figure 6. Yellow spot is center of Tharsis load, at 248.3E, 6.7N (Zuber and Smith, 1997). Polar wander driven by post-Tharsis loads of magnitude much smaller than Tharsis, the subject of this paper, lies close to the green line (Tables 2 and 3), which is within 7° of the great circle perpendicular to Tharsis. Positive loads in the 68E – 248E hemisphere N of the equator, or in the 248E – 68E hemisphere S of the equator, will lever the spin axis toward the top of the figure, relative to the lithosphere (Figure 4). Loads in the 68E –

248E hemisphere N of the equator, or positive loads in the 248E – 68E hemisphere S of the equator, will lever the spin axis toward the base of the figure, relative to the lithosphere (Figure 4).

**Figure 2.** Sketch section showing a) south polar and b) north polar stratigraphy, drawn so that left is up in Figure 1. Principal inferred direction of polar wander is from right to left. Line of section is the green line in Figure 1; each section runs from 70S, through the pole, to 70S in the opposite hemisphere (~2400 km). Locally, the Dorsa Argentea Formation extends beyond 70N.

**Figure 3.** Close-up of region of volcanic loads. a) Uniform load with thickness 100m (total volume 3.0 x $10^6$ $km^3$, of which 77% by volume lies N of equator). Labeled provinces are referred to in Table 1a.; b) 'Small' load (11.2 x $10^6$ $km^3$, 79% ); c) 'Medium' load (19.0 x $10^6$ $km^3$, 77%), d) 'Large' load (35.4 x $10^6$ $km^3$, 79%).

**Figure 4.** North polar projection showing dependence on latitude and longitude of TPW driven by a small positive surface load. Vectors show the direction and magnitude of north pole displacement that would be caused by placing the same load at different geographic locations within the northern hemisphere. Concentric circles are 0N, 30N, and 60 N. Ratio of maximum TPW angles in plots is 1: 1.1 : 34.1. a) Without Tharsis, and with remnant rotational bulge at equator, the amount of TPW is small, and any load displaces the pole along the line of longitude that passes through the load. b) Small Tharsis-driven TPW scenario. The amount of TPW is still small, but in contrast to (a), the direction of TPW generally has a component perpendicular to the meridian that passes through Tharsis. White circle is pre-Tharsis pole; black circle is Tharsis location. c) Large Tharsis-driven TPW scenario. The symbols for Tharsis and the pre-Tharsis pole overlap due to the map projection. TPW is large, but restricted to directions perpendicular to the line of longitude that passes through Tharsis.

**Figure 5.** South polar region showing province-by-province paleopoles (Table 1a). 'Large' load, large TPW scenario. For each province, $T_e$ = 100 km, 200 km and 300 km solutions are at increasing distances from the pole. See Figure 6b for key to polar geologic units.

There is some disagreement about the correct boundary of the Dorsa Argentea Formation. By using the original boundary of Tanaka and Scott (1987), we adhere more closely to the preferred boundary of Ghatan & Head (2004) than to the revised boundary of Tanaka & Kolb (2001).

**Figure 6:** a) North polar region, and b) South polar region, showing:– paleopoles computed from estimates of young volcanic load (triangles; Table 1a for key); paleopoles inferred from geology (nested green circles; Table 1a for key); and distribution of polar and paleopolar deposits (tinted areas). Dark blue line is the convex hull of the Scandia mounds. Green dashed lines define error envelopes based on the deviation of the youngest polar deposits from the current spin axis (2.4±0.3° ≡ 140±20 km, the 'proxy error' discussed in the text). We estimate that the additional error in assigning a centroid locations to a paleopolar deposit ('measurement error') is comparable to or greater than this error, and we show this additional error by the outer green circles. Ice outliers are probably young (contemporaneous with upper Planum Boreum or younger), but are shown because they might be obscuring older deposits.

**Figure 7**. One possible synthesis of Martian volcanic and polar history using TPW constraints and the assumption that areal centroids of polar deposits track the spin axis. Degree of confidence increases with observed offset. The contribution of internal dynamics to TPW is assumed to be small (see §6.3 for a discussion of this assumption). Other less simple syntheses are also consistent with the data. Elevations are not to scale. a) Deposition of Dorsa Argentea Formation (DAF) during Hesperian.

Ongoing volcanism at Elysium (Ely). b) Loading in Quadrants 1 or 3 (e.g. by Elysium or proto-Olympus Mons, Oly; Figure 5) shifts pole by 5.9 degrees in great circle 90 deg from Tharsis in a meridian-ward sense. Deposition of Scandia (Sc) materials intercalated with loading, Basal Unit (BU) postdates loading. c) Further loading in Quadrants 1 or 3 (e.g. by waning Elysium flows, Olympus Mons activity or Arcadia young flood lavas; Figure 5) drives further polar wander of 7.5 degrees in meridian-ward sense. Deposition of Promethei Lingular Lobe of SPLD. d) Loading in Quadrants 2 or 4 (most likely by Tharsis Montes, TM; Figure 5) drives polar wander of 5 degrees in anti-meridian sense. Deposition of Planum Boreum (PB) intercalated with or subsequent to loading. Deposition of the geologically very young Southern Residual Cap (SRC).

**Figure 8.** Magnitude of TPW versus total volume of late Hesperian and Amazonian volcanism, for a range of assumptions about lithospheric thickness and orientation of fossil rotational bulge. Colored bars show range of magnitudes/volumes for which computed TPW matches geologically inferred wander to within ±1°. Red vertical lines are small, medium and large loads. The South Polar Layered Deposits are not plotted because their offset cannot be accommodated by a ‘proportional’ volcanic history.

| a) Combined surface loads | | | Tharsis formed close to equator | | | Tharsis formed far from equator | | |
|---|---|---|---|---|---|---|---|---|
| Load | Volume (x 10^6 km^3 ) | Center of volume | Te = 100km | Te = 200km | Te = 300km (a) | Te = 100km (b) | Te = 200km (c) | Te = 300km (d) |
| 'Small' (A) | 11.2 | 17.4N 220.5E | 89.8N 181.7E | 89.6N 186.1E | 89.5N 185.2E | 83.8N 161.1E | 79.3N 162.3E | 77.3N 163.0E |
| 'Medium' (B) | 19.0 | 15.3N 220.3E | 89.7N 183.4E | 89.4N 181.2E | 89.2N 180.9E | 80.0N 161.8E | 73.0N 162.7E | 69.3N 163.3E |
| 'Large' (C) | 35.4 | 18.0N 217.8E | 89.3N 180.5E | 88.7N 181.3E | 88.4N 180.6E | 71.5N 162.7E | 62.0N 164.9E | 57.7N 165.9E |
| Uniform drape, 100m thickness | 3.0 | 20.8N 214.8E | 89.9N 169.4E | 89.9N 181.2E | 89.9N 177.2E | 88.0N 159.4E | 86.3N 160.7E | 85.4N 160.4E |
| Uniform drape, 1000m thickness | 29.6 | 20.8N 214.8E | 89.4N 175.6E | 88.9N 175.0E | 88.6N 175.2E | 72.5N 162.1E | 61.0N 164.0E | 56.7N 165.0E |
| b) Each province, 'large' load | | | **Tharsis formed close to equator** | | | **Tharsis formed far from equator** | | |
| Load | Volume (x 10^6 km^3 ) | Center of volume | Te = 100km | Te = 200km | Te = 300km (a) | Te = 100km (b) | Te = 200km (c) | Te = 300km (d) |
| Elysium province | 6.6 | 22.3N 152.5E | 89.6N 154.8E | 89.3N 156.1E | 89.1N 156.4E | 76.0N 160.1E | 64.8N 161.2E | 59.3N 162.0E |
| Medusae Fossae province | 2.2 | 1.4S 201.1E | 90.0N 335.4E | 90.0N 305.5E | 90.0N 305.5E | 89.4N 335.4E | 89.4N 335.4E | 89.4N 335.4E |
| Young flood lavas | 4.1 | 33.1N 187.1E | 89.8N 174.3E | 89.7N 167.3E | 89.6N 169.8E | 84.6N 160.3E | 80.3N 160.2E | 77.2N 160.8E |
| Olympus province | 6.1 | 19.4N 224.9E | 89.8N 198.5E | 89.7N 197.2E | 89.6N 196.1E | 85.6N 162.9E | 81.8N 163.4E | 80.4N 163.6E |
| Tharsis Montes province | 10.0 | 3.5N 248.5E | 89.9N 304.4E | 89.9N 304.5E | 89.8N 309.4E | 87.7N 335.0E | 86.0N 334.7E | 85.2N 335.2E |
| Alba province | 4.6 | 41.3N 249.0E | 89.9N 255.2E | 89.8N 252.5E | 89.7N 251.7E | 89.3N 303.0E | 89.1N 287.1E | 89.0N 280.0E |
| Syria | 1.9 | 15.9S | 90.0N | 89.9N | 89.9N | 88.8N | 87.9N | 87.9N |
| province | | 266.5E | 122.5E | 112.7E | 112.7E | 155.3E | 154.2E | 154.2E |

| Crater-retention age | Polar material | Mask (younger materials) | Best fit pole location | (Relative) Confidence that paleopole is due to TPW? |
|---|---|---|---|---|
| **Northern hemisphere** | | | | |
| *Without ice outliers* | | | | |
| 1. Late Amazonian | PB | – | 87.9N 1.4E | Low to Moderate |
| 2. Late Amazonian | OD | PB | 87.6N 181.2E | Moderate |
| 3. Early Amazonian | SC | PB, OD | 85.0N 206.8E | Low |
| 4. Early Amazonian | SC, ST | PB, OD | 81.3N 198.6E | Very Low to Low |
| *With ice outliers* | | | | |
| Late Amazonian? | IO | PB, OD | 84.6N 160.9E | Ice outliers overlie SC and ST, but age relative to PB is uncertain |
| Early Amazonian? | IO, SC | PB ,OD | 83.7N 185.5E | |
| Early Amazonian? | IO, SC, ST | PB, OD | 81.3N 189.7E | |
| **Southern hemisphere** | | | | |
| 5. Upper Amazonian | SRC | – | 87.3S 315E | n.a. (see §1) |
| 6. (?Middle) Amazonian | SPLD | – | 84.3S 162.7E | Moderate |
| 7. Late Hesperian | DAF | SPLD | 82.1S 335.9E (variable cap radius)<br><br>*81.3S 302.1E (fixed cap radius)* | High |

| Unit desig. | Description | Unit thickness or truncation elevation (m) | | | Notes |
|---|---|---|---|---|---|
| | | 'Small' | 'Medium' | 'Large' | |
| *Elysium Province* | | | | | |
| Hhet | Hecates Tholus Fm. | -3000 (T) | -3000 (T) | -3500 (T) | Qualitative backstripping following Tanaka et al. (1992) |
| Ael1 | Elysium Fm., member 1 | 500 | 1000 (T) | -3500 (T) | |
| Ael2 | Elysium Fm., member 2 | 2500 (T) | 2500 (T) | -3500 (T) | |
| AHat | Albor Tholus Fm. | -1000 (T) | -1000 (T) | - 3500 (T) | |
| *Medusae Fossae Formation* | | | | | |
| Combined lower, middle and upper members of Fm. | W lobes | 500 | - 2700 (T) | -2700 (T) | Largely following Bradley et al. (2002). |
| | C lobes | 500 | -2400 (T) | -2800 (T) | |
| | E lobes | 500 | -2000 (T) | -3000 (T) | |
| | Far E lobe | 500 | +500 (T) | -500 (T) | |
| *Young flood lavas* | | | | | |
| Ael3-4 | Elysium Fm., members 3-4 | 100 | 200 | 500 | 'Small' values from SHARAD and MARSIS subsurface contacts |
| Achu | Younger channel system material, undivided | 75 | 300 | 1000 | |
| Aa1 | Arcadia Fm., member 1 | 150 | 200 | 500 | Not between -55 and +10 deg E |
| Aa2 | Arcadia Fm., member 2 | 50 | 100 | 200 | 'Small' values from SHARAD and MARSIS subsurface contacts |
| Aa3 | Arcadia Fm., member 3 | 70 | 100 | 500 | |
| Aa4 | Arcadia Fm., member 4 | 30 | 100 | 200 | |
| Aa5 | Arcadia Fm., member 5 | 30 | 100 | 200 | Only between -180 and -150 deg E |
| Achp | Younger | 30 | 30 | 100 | 'Small' value from |
| | flood-plain material | | | | SHARAD and MARSIS subsurface contacts |
| *Olympus Mons province* | | | | | |
| Aoa1 | Olympus Mons Fm., aureole member 1 | 1000 | 1000 | 1000 | |
| Aoa2-Aoa4 | Olympus Mons Fm., aureole member 2 – member 4 | 1500 | 1500 | 2200 | |
| Aos | Olympus Mons Fm., shield member | 3000 | +0 (T) | - 2000 (T) | |
| Aop | Olympus Mons Fm., plains member | 500 | 1500 | 2500 | |
| As | Slide material | 3000 | +0 (T) | - 2000 (T) | Only between -128 E and -140 E. As Aos |
| Ae | Eolian deposits | 1500 | 1500 | 2200 | Only between -130 and -150E. As Aoa2-4. |
| *Tharsis Montes province* | | | | | |
| AHt3 + At6 + As | Arsia | + 8000 (T) | + 6000 (T) | + 4000 (T) | |
| | Pavonis | + 5000 (T) | + 4000 (T) | + 2000 (T) | |
| | Ascraeus | + 4000 (T) | + 2500 (T) | + 0 (T) | |
| | “marginal” | 100 | 250 | 250 | |
| At5 | Tharsis Montes Fm., member 5 | 200 | 500 | 500 | |
| At4 (distal) | Tharsis Montes Fm., member 4 | 100 | 250 | 500 | |
| *Alba Patera province* | | | | | |
| Aau | Alba Patera Fm., upper member | +4500 (T) | +2000 (T) | +0 (T) | Late Hesperian apron and Earliest Amazonian summit according to |
| Aam | Alba Patera Fm., middle member | 1000 | +2000 (T) | - 1000 (T) | Ivanov and Head (2006) |
| Hal | Alba Patera Fm., lower member | 100 | 200 | 500 | |
| *Syria province* | | | | | |
| Hsl | Syria Planum Fm., lower member | 200 | 500 | 1000 | |
| Hsu | Syria Planum Fm., upper member | 200 | 500 | 1800 | |
| *Other* | | | | | |
| AHcf | Ceraunius Fossae Formation | 200 | 450 | 1000 | |
| "Nf" | Highly-deformed terrain materials | 200 | 500 | 2500 | Adjacent to AHcf only |
| Ht1-2 | Tharsis Montes Fm., members 1-2 | 100 | 200 | 500 | |
| Apk | Knobby plains material | 100 | 200 | 500 | Only near to Tharsis Montes |

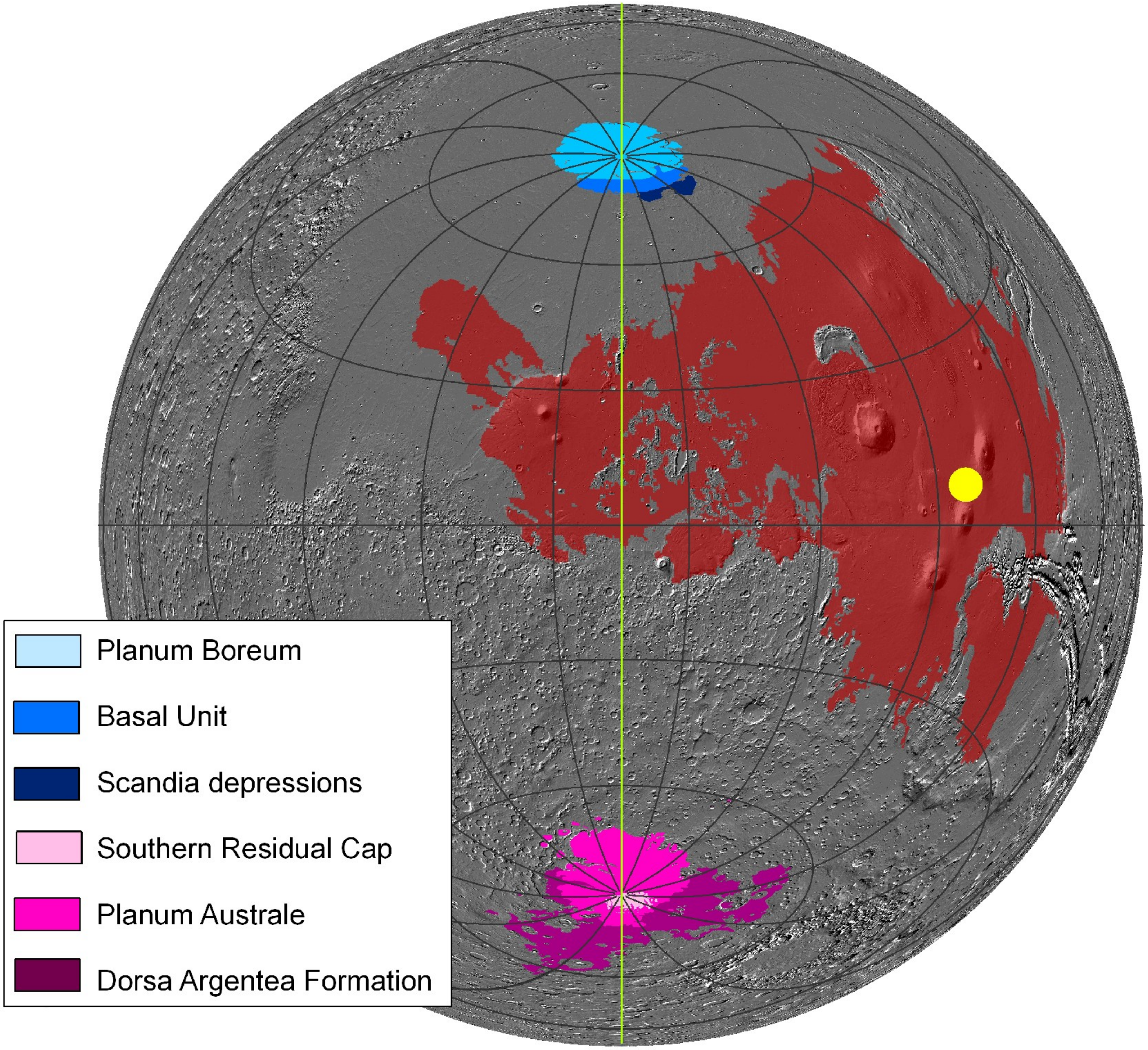

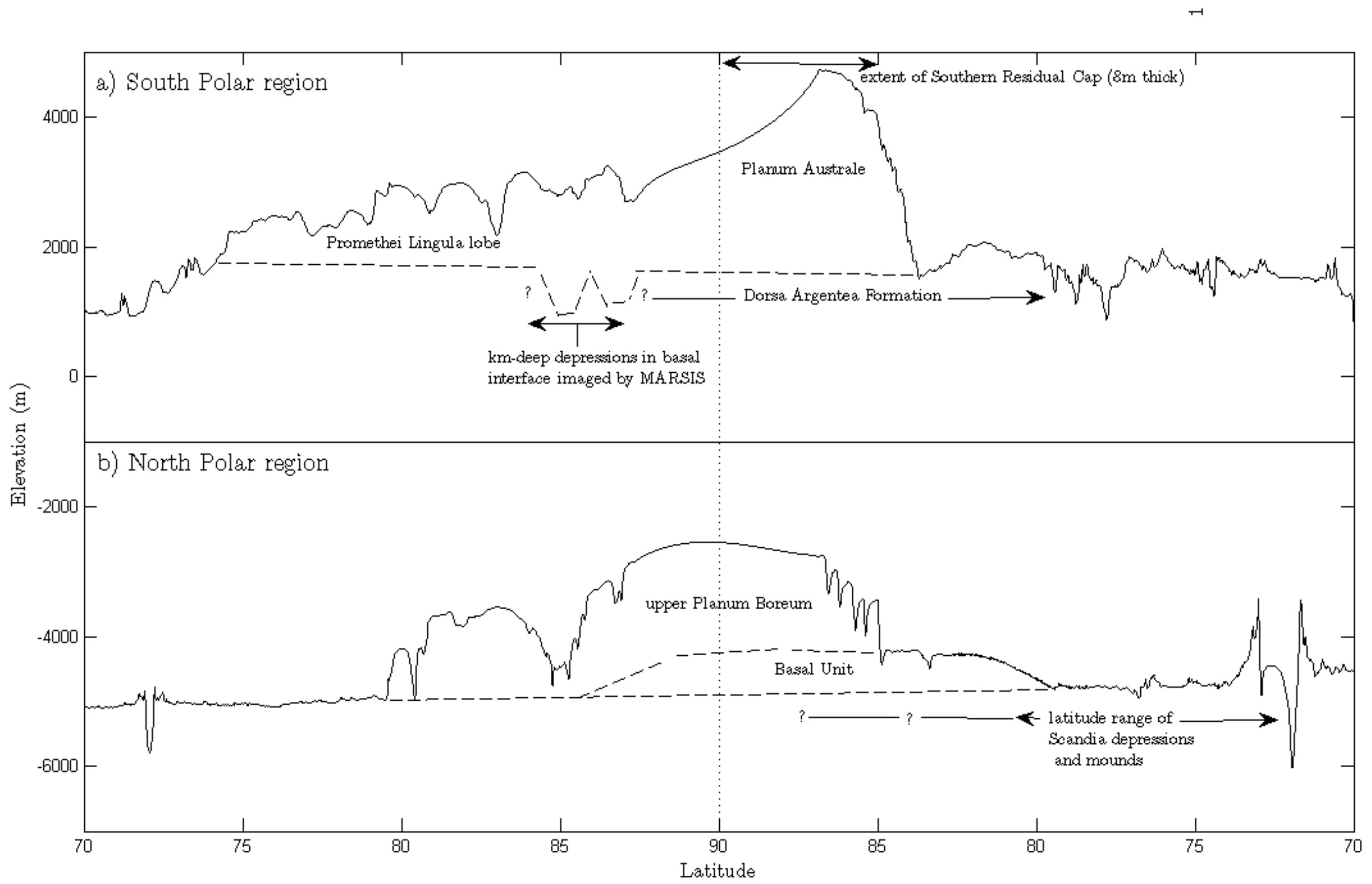
a) South Polar region
extent of Southern Residual Cap (8m thick)
Planum Australe
Promethei Lingula lobe
Dorsa Argentea Formation
km-deep depressions in basal
interface imaged by MARSIS
b) North Polar region
upper Planum Boreum
Basal Unit
latitude range of
Scandia depressions
and mounds
Elevation (m)
Latitude

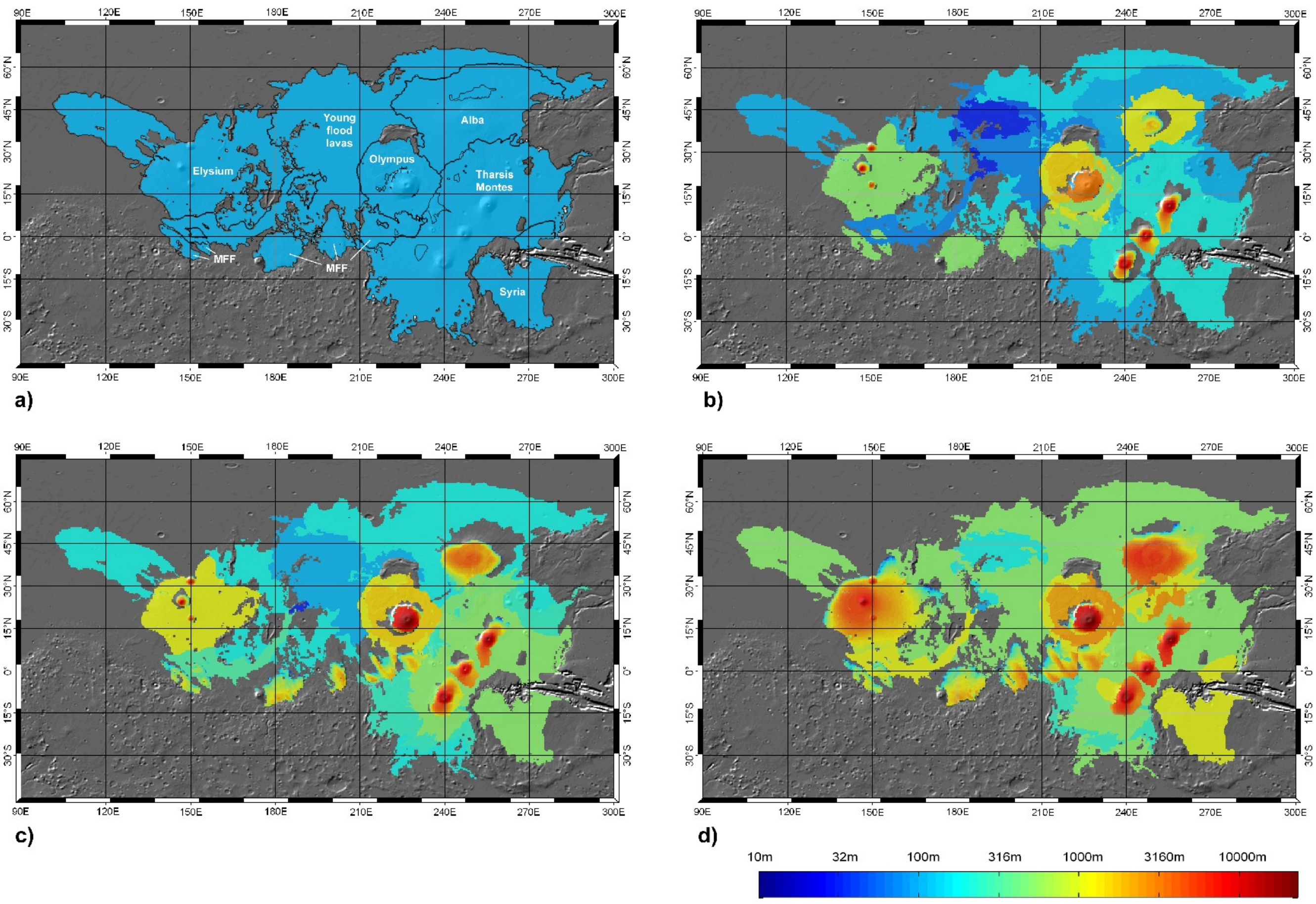

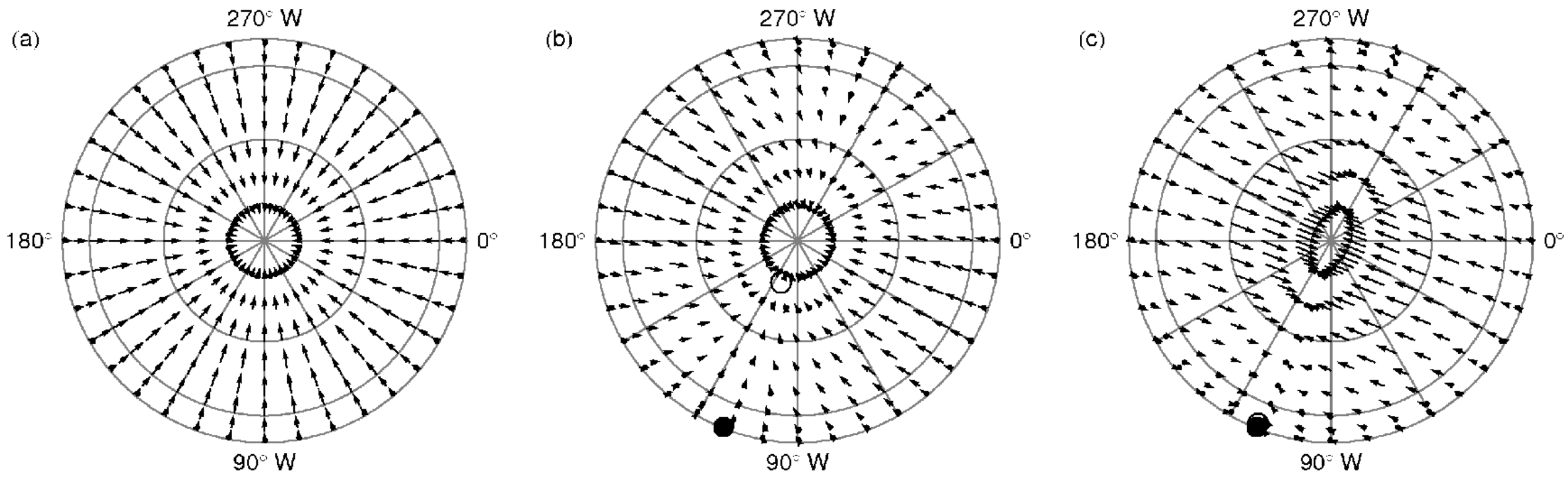

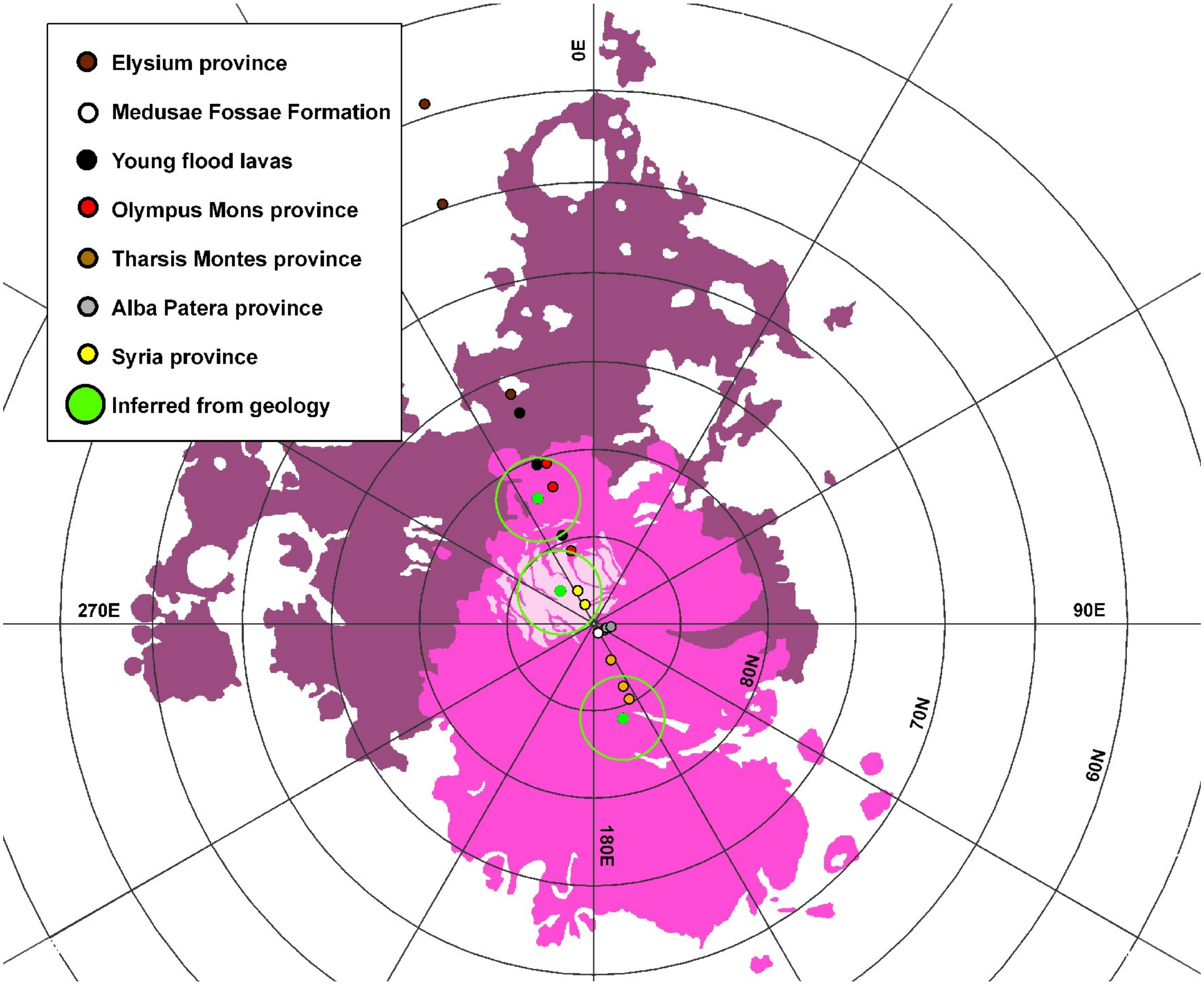

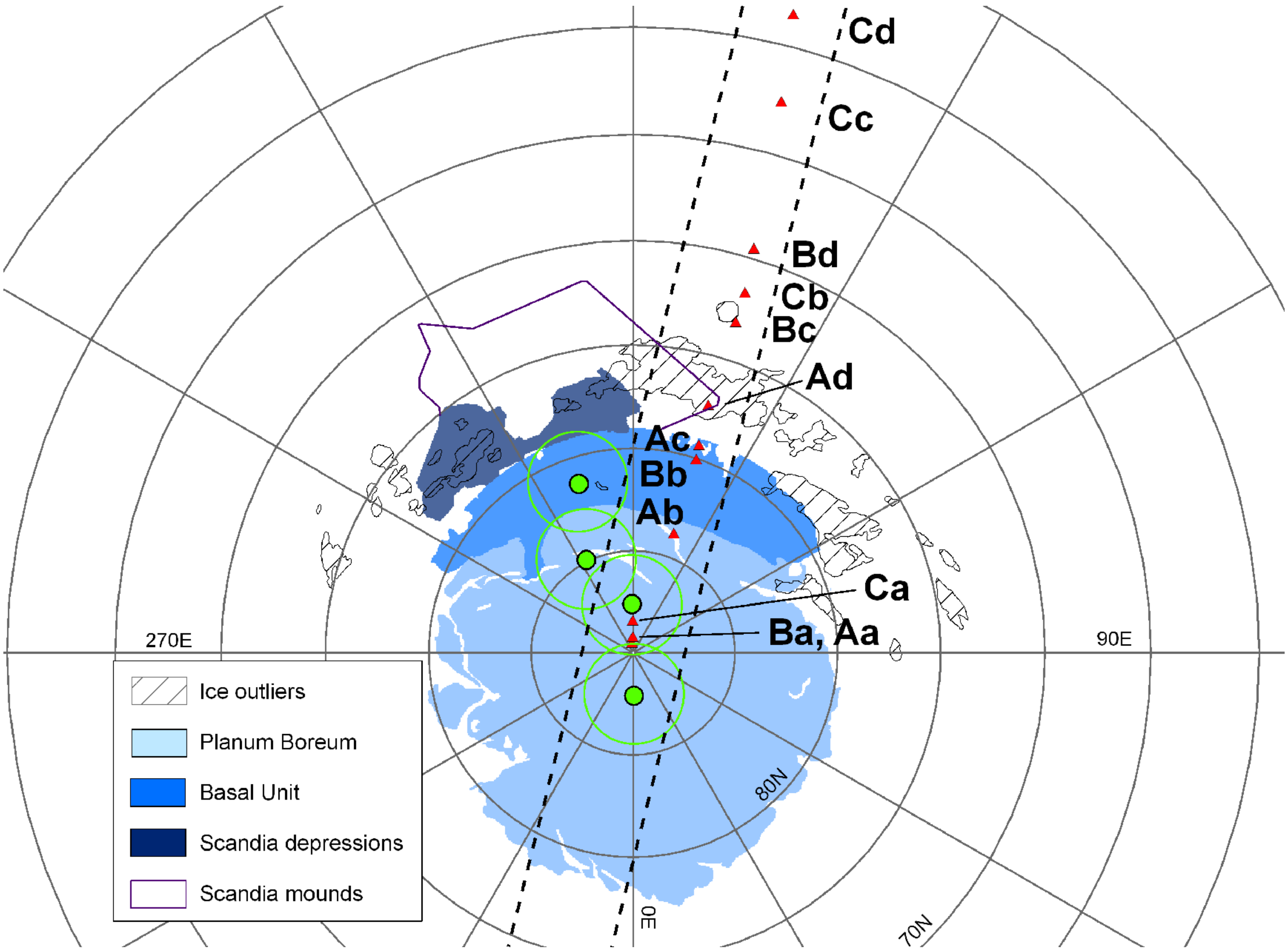

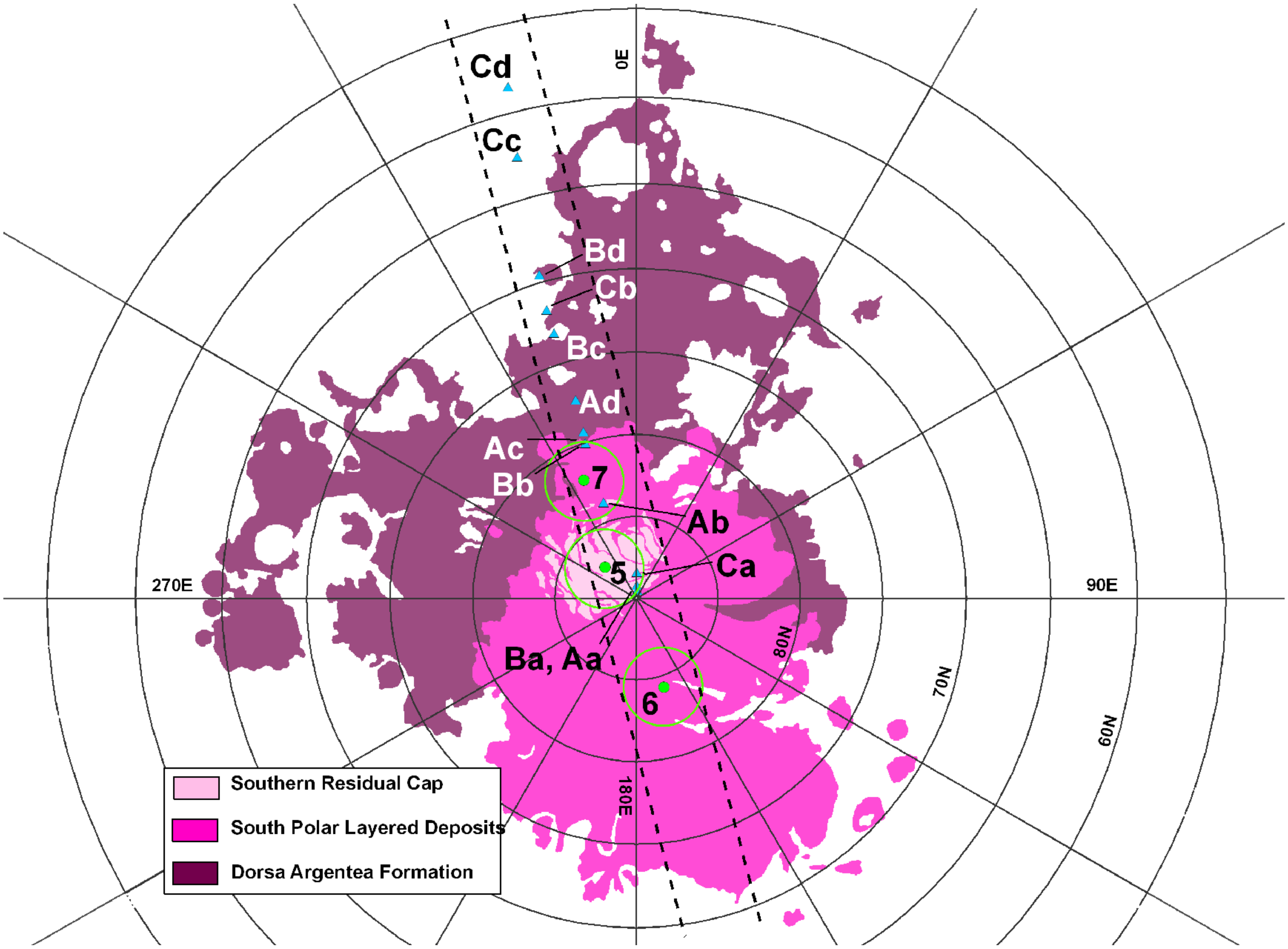

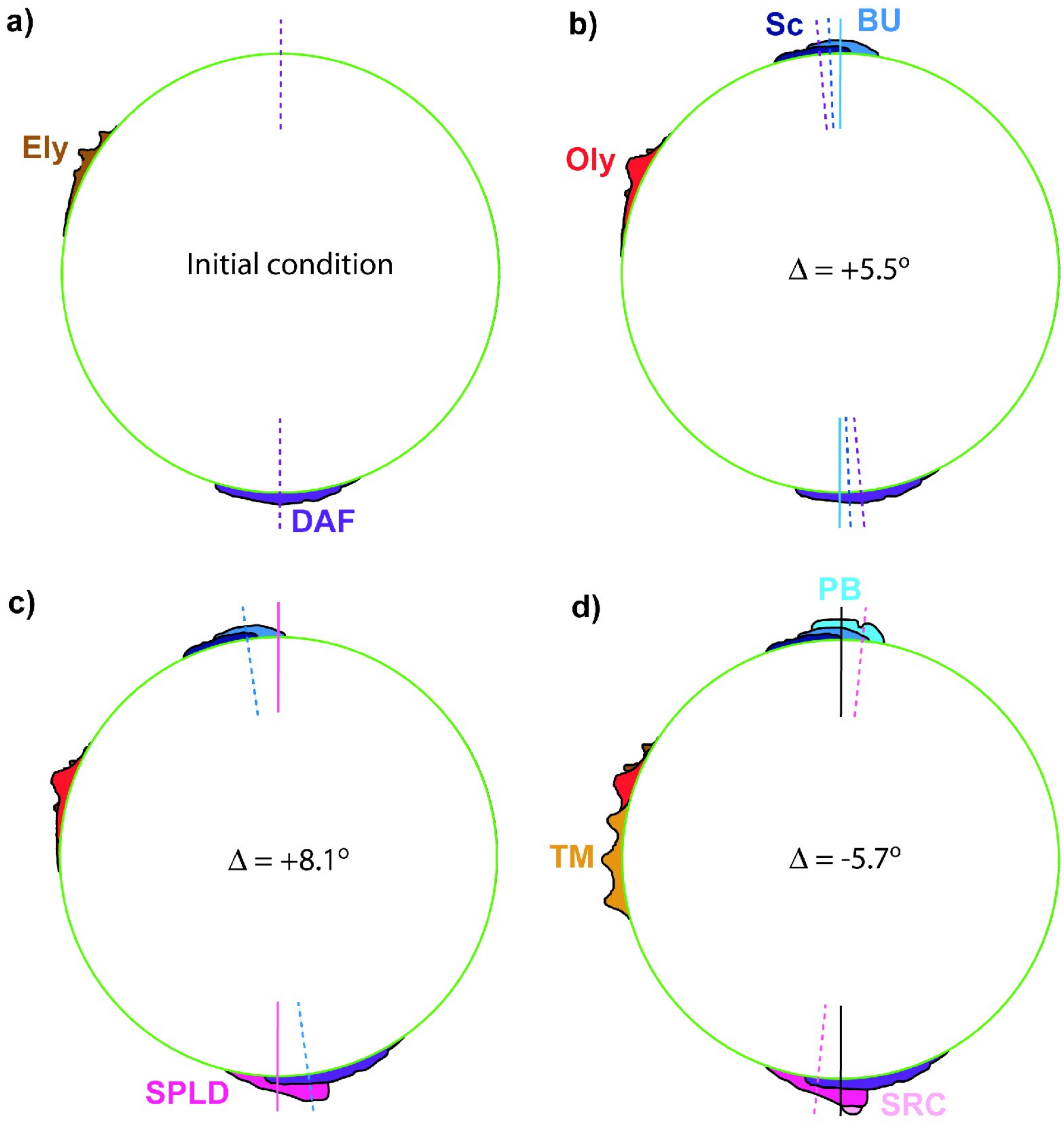

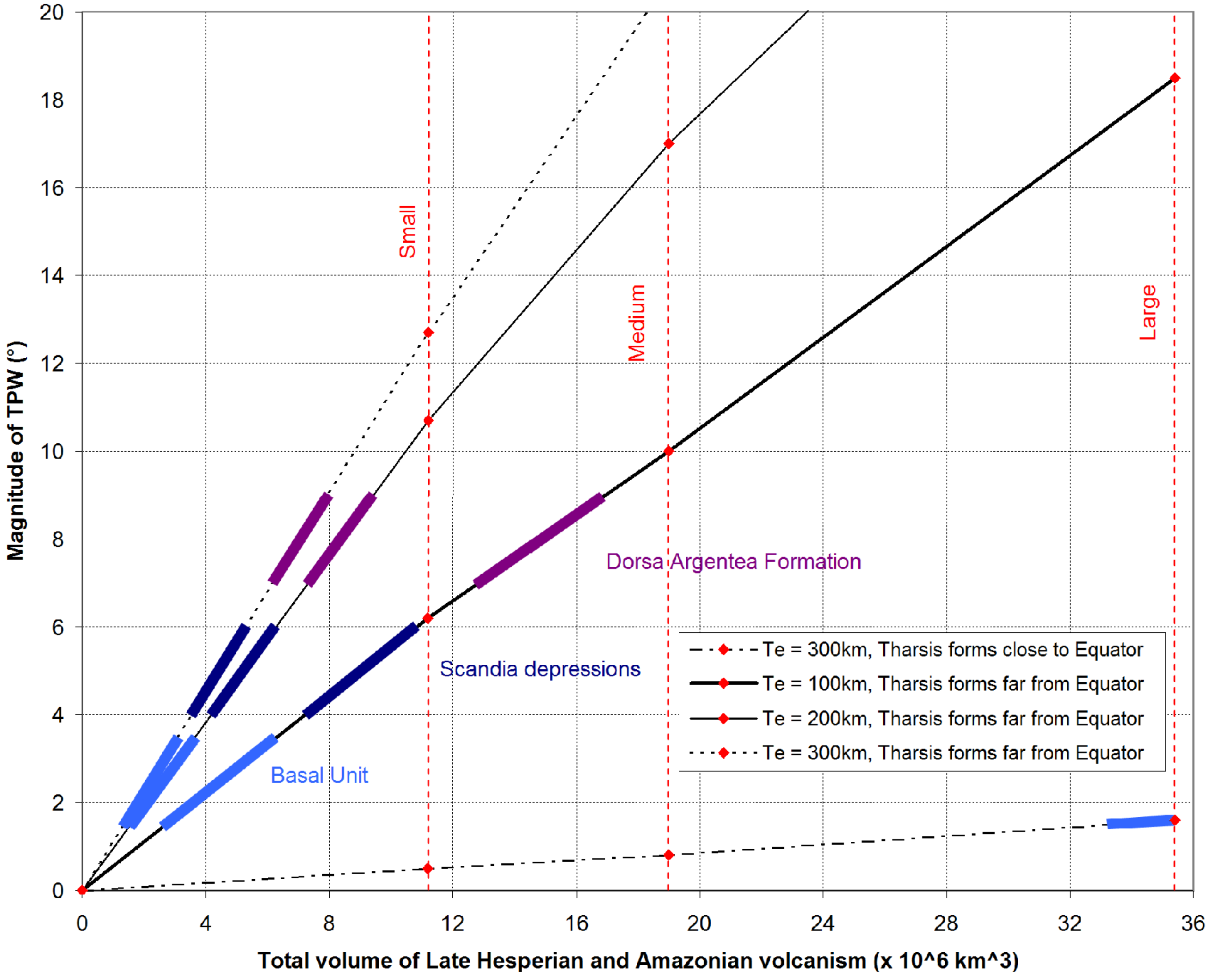

Magnitude of TPW (°)
Total volume of Late Hesperian and Amazonian volcanism (x 10^6 km^3)
Small
Medium
Large
Dorsa Argentea Formation
Scandia depressions
Basal Unit
Te = 300km, Tharsis forms close to Equator
Te = 100km, Tharsis forms far from Equator
Te = 200km, Tharsis forms far from Equator
Te = 300km, Tharsis forms far from Equator